# Nanosecond Ferroelectric Switching of Intralayer Excitons in Bilayer 3R-MoS$_2$ through Coulomb Engineering

Jing Liang[1,2]#, Yuan Xie[1,2]#, Dongyang Yang[1,2]#, Shangyi Guo[1,2], Kenji Watanabe[3], Takashi Taniguchi[4], Jerry I. Dadap[1,2], David Jones[1,2], Ziliang Ye[1,2]*

[1] Quantum Matter Institute, The University of British Columbia, Vancouver, BC V6T 1Z4, Canada

[2] Department of Physics and Astronomy, The University of British Columbia, Vancouver, BC V6T 1Z1, Canada

[3] Research Center for Functional Materials, National Institute for Materials Science, 1-1 Namiki, Tsukuba 305-0044, Japan

[4] International Center for Materials Nanoarchitectonics, National Institute for Materials Science, 1-1 Namiki, Tsukuba 305-0044, Japan

# These authors contributed equally to this manuscript.

* Correspondence: zlye@phas.ubc.ca

**Abstract**

High-speed, non-volatile tunability is critical for advancing reconfigurable photonic devices used in neuromorphic information processing, sensing, and communication. Despite significant progress in developing phase change and ferroelectric materials, achieving highly efficient, reversible, rapid switching of optical properties has remained a challenge. Recently, sliding ferroelectricity has been discovered in 2D semiconductors, which also host strong excitonic effects. Here, we demonstrate that these materials enable nanosecond ferroelectric switching in the complex refractive index, largely impacting their linear optical responses. The maximum index modulation reaches about 4, resulting in a relative reflectance change exceeding 85%. Both on and off switching occurs within 2.5 nanoseconds, with switching energy at femtojoule levels. The switching mechanism is driven by tuning the excitonic peak splitting of a rhombohedral molybdenum disulfide bilayer in an engineered Coulomb screening environment. This new switching mechanism establishes a new direction for developing high-speed, non-volatile optical memories and highly efficient, compact reconfigurable photonic devices. Additionally, the demonstrated imaging technique offers a rapid method to characterize domains and domain walls in 2D semiconductors with rhombohedral stacking.

**Main**

Switchable optical materials can enable non-volatile tunability in photonic devices with applications that include neuromorphic computing and artificial intelligence, quantum information processing, optical communications, and optical sensing[1-6]. Unlike volatile tuning, no power is needed to maintain a switched state in a non-volatile device, which can therefore serve as an optical memory/memristor[7,8]. To this end, much effort has been focused on the large refractive index tuning in chalcogenide-based phase change materials, which requires pico- to nano-joules of thermal energy to switch[9,10]. Recently, barium titanate thin films have been developed as a promising ferroelectric material to achieve a switchable Pockels effect with record-low switching energy and a refractive index modulation up to $10^{-3}$ under a constant electric field[11]. Here we present a new scheme for non-volatile switching of the complex refractive index associated with the excitonic resonances in two-dimensional (2D) semiconductors. The modulation in both the refractive index and extinction coefficient is about 4, resulting in a relative reflectance change exceeding 85%. Since no heating is involved in our switching mechanism, the switching energy can be reduced to femtojoule levels, with both on and off switching occurring in less than 2.5 ns, making it the fastest non-volatile tunability reported for optical materials[7]. Our scheme relies on the dynamic tuning of the Coulomb screening of strong excitonic effects in 2D transition metal dichalcogenides (TMDs) that are rhombohedrally stacked to enable sliding ferroelectricity.

Sliding ferroelectricity is a hysteretic phenomenon in 2D van der Waals materials with specific stacking orders, where an electric field induces one layer of the material to slide relative to the other due to an interfacial polarization arising from the interlayer coupling[12-15]. This effect can occur in traditionally non-ferroelectric materials and has been electrically probed,[16-21] with atomic structure changes confirmed by scanning probe microscopy[22] and electron microscopy[23,24]. Initially found in artificially stacked 2D materials with marginal twists, interfacial polarization and its switching have recently also been observed in chemically synthesized rhombohedral (3R) TMDs[25-30], where preexisting domain walls play a key role in lowering the coercive field. Since atomically thin TMDs are semiconductors with much reduced Coulomb screening and extraordinary excitonic effects[31-34], the switch in the stacking configuration can also be reflected in the excitonic properties as optical contrasts, which have been observed between intermediate states in thick layers with multiple interfaces[26,35]. Nevertheless, in an intrinsic 3R bilayer with only one interface, the stable stackings before and after the switch form a mirror image pair (i.e., AB and BA), rendering them indistinguishable with conventional linear optical probes. Optically, the AB and BA stackings have therefore only been resolved by photoluminescence spectroscopy under an external electric field[26,36], which are incompatible with the applications discussed above.

On the other hand, the mirror symmetry between AB and BA stacking configurations can be broken by introducing an asymmetric dielectric environment. Previous work has shown that the Coulomb interaction in atomically thin TMDs can be modified by engineering the local environment[37-40]. Large Coulomb screening can significantly reduce the quasiparticle bandgap and exciton binding energy, and such a screening effect has a distance sensitivity down to a single monolayer[37]. Consequently, when a TMD bilayer is placed on a substrate, the lower layer will experience greater screening than the upper layer that is not in direct contact with the dielectric substrate (Fig. 1a). If the intrinsic excitonic responses in two layers are identical, such asymmetric screening will result in no net contrast between the two stackings. However, the interlayer coupling in a bilayer 3R-TMD naturally breaks the layer symmetry, enabling significant non-volatile tuning of the complex refractive index near the excitonic resonance.

**Stacking-dependent excitonic resonances**

In a rhombohedral $MoS_2$ bilayer, the two layers experience different chemical environments and can have distinct exciton peaks[36,41]. Here we define the A layer as having the molybdenum atom aligned with the sulfide atom in the B layer (Fig. 1b). Under this definition, the molybdenum atom in the B layer is not aligned with the sulfide atom in the A layer. Consequently, the lowest excitonic resonance in the monolayer $MoS_2$ splits into two peaks in a rhombohedral bilayer, each peak originating from an intralayer exciton composed of an electron and a hole located within the same layer. We have previously confirmed that the high-energy peak of this pair originates from the K/K' point transition in the A layer ($X_A$) while the low-energy peak arises from the B layer ($X_B$)[41]. In a symmetric dielectric environment, the energy splitting between the two exciton peaks, $\delta_0$, is determined by the interlayer coupling strength, independent of the stacking configuration. However, this degeneracy is lifted when the dielectric environments of the upper and lower layers differ. For example, Coulomb screening from a substrate, as depicted in Fig. 1b, modifies the energy splitting. In one stacking configuration where the B layer is at the bottom of the A layer and is on top of the substrate, the low-energy $X_B$ experiences a redshift ($\Delta$) and the peak splitting increases ($\delta_{AB} = \delta_0 + \Delta$). Conversely, when the A layer is at the bottom of the B layer, the high-energy $X_A$ experiences a redshift, resulting in a decrease in energy splitting ($\delta_{BA} = \delta_0 - \Delta$). As a result, when the stacking is switched through sliding, the energy splitting between the two excitons changes accordingly. In real materials, the top layer may also experience some degree of screening from the substrate, but the peak separation contrast between the two stackings should persist.

Following this picture, we studied a 3R-MoS$_2$ bilayer directly exfoliated onto a SiO$_2$/Si substrate. The coexistence of both stacking configurations is reflected in the surface potential contrast, mapped by electrostatic force microscopy (EFM) (Fig. 2a). Due to the polarization-induced interlayer potential, the surface potential in the AB domain is higher than that in the BA domain[42]. Subsequently, we conducted optical reflectance contrast (RC) spectroscopy at 1.6 K to investigate the excitonic responses of the two differently stacked domains, marked by orange and blue dots. Fig. 2b presents the RC spectra and their second energy derivatives taken at these two locations, respectively. The AB domain clearly exhibits two distinguishable peaks while the BA domain displays an effective single peak due to the unresolved peak splitting, as illustrated in the lower right panel of Fig. 1b.

To quantify the asymmetric screening effect, we performed a mapping of the RC across the sample (Fig. 2c). Although optical imaging has lower resolution due to the diffraction limit, the reflectance at 1.919-eV photon energy shows a contrast similar to the EFM map. The statistics of the excitonic peak splitting among AB domains yield an average of $\delta_{AB} = 35 \pm 3$ meV (Figs. 2d, e). In a separate experiment, we fabricated a switchable 3R-MoS$_2$ bilayer device in a symmetric dielectric environment (Fig. S1). The excitonic peak splitting in both stackings is about 15 meV, which is the intrinsic energy splitting ($\delta_0$) of the rhombohedral bilayer. Consequently, we conclude that the extra redshift $\Delta$ caused by the SiO$_2$ substrate is approximately 20 meV. This value suggests $\delta_{BA}$ is about 5 meV smaller than the excitonic linewidth in our unencapsulated sample (~40 meV), which explains the unresolved peak splitting in Fig. 2b. We found that the excitonic contrast induced by the asymmetric Coulomb screening persists up to room temperature, enabling the direct visualization of the AB and BA domains using an optical microscope with a suitable band pass filter tuned to the excitonic resonance energy (Fig. 2f and S2). Such an optical contrast can be used to rapidly screen the domain structure in chemically synthesized films on a wafer scale, benefiting the fasting growing 3R-TMD crystal growth community[43,44].

**Dynamic tuning of excitonic peak splitting**

Next, we show that the excitonic peak splitting can be dynamically tuned in a device that can switch the stacking configuration through sliding ferroelectricity (Fig. 3a). The device consists of a MoS$_2$ bilayer, with the lower layer adjacent to a few-layer graphene (FLG) and the upper layer next to an *h*-BN layer, forming an asymmetric Coulomb screening environment (see the optical image of the device in Fig. S3). An additional FLG is placed on top to serve as an electrode for applying a vertical electric field. The entire stack is placed on a SiO$_2$/Si substrate. The bilayer contains a preexisting domain wall (DW), and as we have previously demonstrated, an out-of-plane electric field can induce an in-plane pressure on this DW since the neighboring domains have opposite interfacial polarizations[26]. When the induced free energy difference across the DW exceeds the local pinning potential, the DW can be released and sweeps through a large area until trapped by another pinning center. This stacking configuration switching is analogous to polarization poling in conventional ferroelectrics, except that the generation of DWs is challenging in sliding ferroelectrics and most switching relies on the motion of preexisting DWs[26]. Since excitonic peak splitting is enhanced in one stacking and reduced in the other, we can optically probe the stacking configuration by measuring the RC spectrum after electrical poling. Here the RC spectroscopy measurements are used to characterize the steady configurations. The switching dynamics are measured using a single wavelength laser resonant with the exciton energy as discussed in the next section.

The electric poling field ($E_p$) dependent RC spectra are measured at the center of the device using a broadband white light source focused down to a micron-sized spot (Fig. 3b). When the poling field is scanned from positive to negative, an unresolved single peak at ~1.909 eV is observed at the positive limit, indicating the initial stacking is BA. When $E_p$ exceeds the coercive field in the negative direction ($E_c^- = -0.042$ V/nm), a transition in the spectrum occurs, revealing two excitonic peaks with a 22-meV splitting. Such a transition is consistent with the substrate-only experiment and indicates that the stacking configuration has switched to AB. The energy splitting differences between the two types of asymmetric dielectric environments are attributed to variations in Coulomb screening, which depend on both the absolute dielectric constant values and the contrasts between those of the substrate and superstrate[37]. Furthermore, polarization-switching-induced free carrier density changes in the FLG layer are estimated to have a negligible impact ($\sim$ 1 meV, see details in the Supplementary Information) on $\delta_{AB}$ and $\delta_{BA}$[45]. In contrast, no spectral change is observed in the device with symmetric hBN encapsulation (Fig. S1). Similarly, a forward $E_p$ scan shows an opposite stacking switch from AB to BA when the external field exceeds the positive coercive field ($E_c^+ = 0.038$ V/nm). The coercive fields are not symmetric since they are mostly determined by the trapping potential of the initial and final pinning centers. As the reflectance mapping below will show, the intermediate state is caused by the pinning centers within the focus spot, which can trap DWs under a small poling field with variations across devices (Fig. S4-S5). These intermediate states can be engineered for multi-bit optical memristors in the future[7]. Compared to the case of the SiO$_2$ substrate, $\delta_{AB}$ is smaller in the encapsulated device because the peak separation is determined by the relative screening strength between the substrate and the capping layer.

At the photon energy of 1.909 eV, which is near the center of the unresolved peak and the middle of the double peak, a large reflectance contrast is observed before and after the switch. The BA state yields a reflectance, $R_{BA}$, of 8.56%, which is over six times higher than that of the AB state ($R_{AB}$=1.23%), corresponding to a relative change, $(R_{BA}-R_{AB})/R_{BA}$, of 86%. (The unnormalized difference spectrum is plotted in Fig. S6). The switch is non-volatile with a clear hysteretic dependence on the poling field (Fig. 3c). Given the device geometry and the observed coercive field, the electrostatic energy needed to switch the stacking in our nonoptimized device is less than one picojoule, indicating its potential for reconfigurable photonic applications. The switching energy can be further lowered by reducing the device capacitance in future designs.

Additionally, we retrieved the complex refractive index change in the 3R-MoS$_2$ bilayer by fitting the reflectance spectra in the AB and BA stackings (Fig. 3d). The spectra near the excitonic resonance are fitted with a dielectric function model composed of two Lorentz peaks and a background dielectric constant, while the multilayered environment is taken into account via a transfer-matrix model[46-48] (see details in the Supplementary Information). The retrieved refractive index ($n$) and extinction coefficient ($k$) are plotted in Fig. 3e. Near the excitonic resonance, a maximum modulation of about 4 is observed in both optical constants (Fig. S6). The sum of the two oscillator strengths remains nearly constant during the switching event. To explore the compatibility of our scheme with various applications, we also investigated the switchability of the device at room temperature (Fig. 3f and S7), where the peak splitting is no longer resolvable due to thermal broadening. Since the merged peaks have very different effective linewidths, a sharp change can still be observed in the reflectance spectra when the poling field exceeds the coercive field. At room temperature, the reflectance changes from 1.42% to 1.95% at the resonance peak, resulting in a relative change of 27%. With the light currently propagating vertically through a 1.4-nm thin active layer, we expect the modulation depth can be further improved by integrating the material into an optimized photonic device. Recent studies have shown that sliding ferroelectricity-based memory can be retained for months and endure over $10^{11}$ switching cycles[20,21], making it suitable for most reconfigurable photonic devices.

**Switching speed**

The large reflectance contrast between the AB and BA stackings can also reveal how the domain and DW evolve with the poling field (Fig. 4a). We focused a 648.2-nm continuous wave laser on a diffraction-limited spot on the device and measured its reflected intensity using a high-speed avalanche photodiode (APD) while the sample is raster-scanned. Initially, the device is poled with a large negative field and most areas exhibited low reflectivity, indicating a uniform AB stacking, except for some highly reflective regions attributed to bubbles, wrinkles, and flake edges (see the optical image in Fig. S3). After a poling voltage of 0.55 V is applied to the device, a small BA domain with higher reflectance emerged near the edge of the FLG electrode. Since it is very challenging to generate new DWs at such a low field, we attribute the new domain to the release and leftward movement of a preexisting DW pinned at the edge. As the poling field increased, the BA domain expanded further, eventually reaching the bubble region. Finally, at 0.9 V, the DW moved downward and stopped at the lowest point, converting half of the sample into a homogeneous BA domain. The spectra in Fig. 3, measured in the middle of this switched BA domain (marked in Fig. 4a and S3), reflects the propagation of this single domain wall. When a negative poling field exceeding the negative coercive field was applied, the BA domain shrank as the DW returned to its initial edge. We note that all states remained non-volatile as the reflectance mapping was carried out with the electric poling field turned off.

Finally, we performed the first real-time optical measurement of the switching dynamics in sliding ferroelectrics. We continuously monitored the reflected laser beam intensity (top blue panel in Fig. 4b) while applying a bipolar square wave voltage to the device through an arbitrary waveform generator (AWG) (bottom red panel). The measurement spot is the same as marked in Fig. 4a. The electric field level was chosen to be well above the coercive field but small enough to keep the Fermi level in the $MoS_2$ bandgap, so that the doping effect is minimized in such a single-gate device (Fig. S8). In addition, since the electric field used in the speed measurement does not match the resonance condition, interlayer excitons do not affect the transient measurements (Fig. S8). Although the square wave voltage is not optimized due to a limited AWG bandwidth as well as impedance mismatch, the two signals displayed clear correlation with distinct high/low contrast. The measured dynamic contrast is similar to the contrast between the steady on and off states at zero electric field, confirming a negligible contribution from the field-induced doping. Both rise and fall times in the optical reflectance signal are about 2.5 ns (shown in the insets of Fig. 4b), limited by the bandwidth of the APD and oscilloscope employed (SI). Interestingly, the step in the optical reflectance is significantly shorter and cleaner than that in the electric field, indicating that the abrupt change was caused by a fast-moving DW released by an electric field that exceeds the pinning threshold, and the duration spent by the DW within the probing spot was very short. The intrinsic switching speed is thus determined only by how quickly the DW sweeps across the focus spot. Our real-time measurement results therefore set a lower bound on the DW velocity of approximately 500 m/s. Assuming a free DW can propagate at the speed of sound[20,21], the intrinsic switching time for a diffraction-limited focus spot is expected to be on the order of 100s of ps.

In summary, we have shown that nanosecond ferroelectric tuning of the complex refractive index with a magnitude of approximately four is achievable near the excitonic resonance of a 3R-$MoS_2$ bilayer by introducing an asymmetric Coulomb screening environment. Without further optimization of the photonic environment, a large non-volatile reflectance change was observed in our device with a high energy efficiency. Compared to the contrast observed between intermediate states in thick layers with multiple interfaces[26,35], the engineered optical contrast between AB and BA stackings in the bilayer with a single interface is larger and more robust. In the future, one can improve the scalability of the device by engineering the domain walls and pinning center in chemically synthesized films. The integration of this phenomenon into photonic waveguides can also bring about new functionalities for integrated photonics. Finally, our optical imaging method enables the exploration of domain wall dynamics and rapid characterization of the domain distribution in 3R-TMDs with a high throughput.

## Acknowledgement

We acknowledge support from the Natural Sciences and Engineering Research Council of Canada, Canada Foundation for Innovation, New Frontiers in Research Fund, Canada First Research Excellence Fund, and Max Planck–UBC–UTokyo Centre for Quantum Materials. Z.Y. is also supported by the Canada Research Chairs Program. K.W. and T.T. acknowledge support from JSPS KAKENHI (Grant Numbers 19H05790, 20H00354 and 21H05233).

## Author Contributions

Z.Y. conceived and supervised the project. J.L., S.G., Y.X., D.Y., K.W. and T.T. prepared the materials. J.L. conducted the EFM characterization. J.L. and S.G. performed the optical spectroscopy in bare flakes. J.L. and Y.X. fabricated the switchable device and performed the measurements with assistance from D.Y. and D.J.. Data analysis was carried out by J.L., Y.X., D.Y., and Z.Y.. J.L., J.D., D.J., and Z.Y. wrote the manuscript with input from all co-authors. J.L., Y.X., and D.Y. contributed to this work equally.

**Competing interests**

The authors declare no competing interests.

**Figures and captions:**

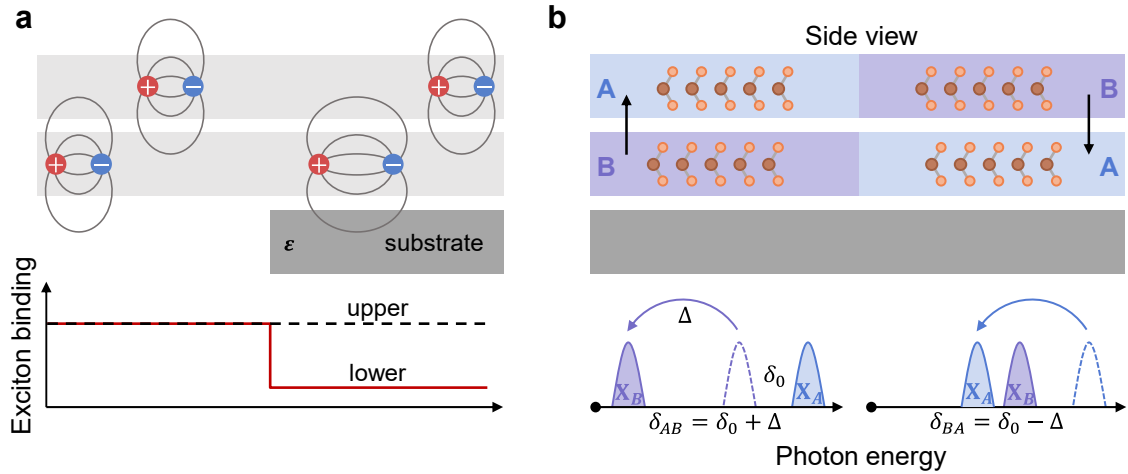

**Figure 1 │ Tuning the excitonic peak splitting through asymmetric dielectric screening.** (**a**) Schematic illustration of the screening effect of a dielectric substrate on the intralayer exciton in the lower layer of a TMD bilayer. The exciton binding energy is lowered due to the reduced Coulomb interaction. (**b**) Schematic of the stacking-dependent excitonic peak splitting in a 3R-MoS$_2$ bilayer on a dielectric substrate. Black arrows indicate the direction of spontaneous polarization. The left and right halves illustrate the AB and BA stackings, respectively. The intrinsic peak splitting and screening-induced redshift are denoted by $\delta_0$ and $\Delta$, respectively.

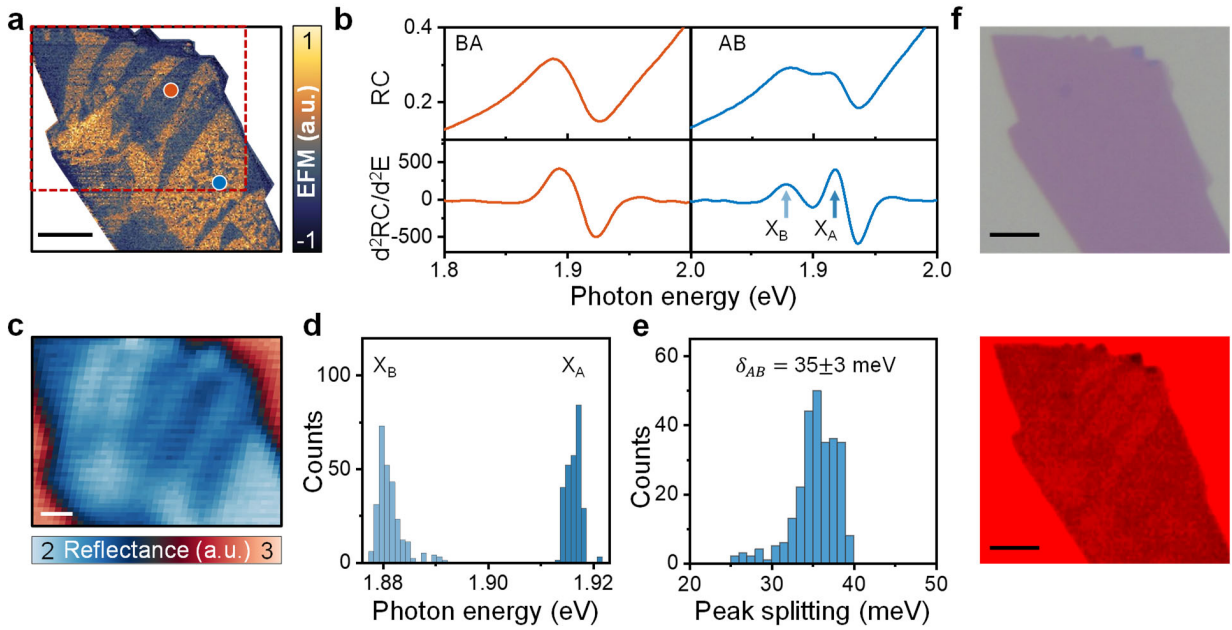

**Figure 2 | Excitonic contrast between AB and BA domains. (a)** Electrostatic force microscopy (EFM) map of a 3R-MoS$_2$ bilayer with mixed AB and BA domains. The flake is directly exfoliated onto a SiO$_2$/Si substrate. The EFM contrast is caused by the surface potential difference between AB and BA stackings. Scale bar: 4 μm. **(b)** Reflectance contrast (RC) spectra and their second energy derivatives for the BA and AB domains. The measurement spots are marked by orange and blue dots in panel a. Two distinct exciton peaks are observed in the AB domain (X$_A$ and X$_B$), while the BA domain exhibits only one unresolved peak. Optical measurements are taken at 1.6 K. **(c)** Optical reflectance mapping (hν =1.919 eV) shows a similar contrast as in the red box in panel a. Scale bar: 2 μm. **(d, e)** Peak energy statistics of X$_A$ and X$_B$ indicate a splitting energy of 35 ± 3 meV. **(f)** Optical images of the 3R-MoS$_2$ bilayer captured without any filter (top), and with a band-pass filter centered at 670 nm with a bandwidth of 10 nm (bottom). Scale bar: 4 μm.

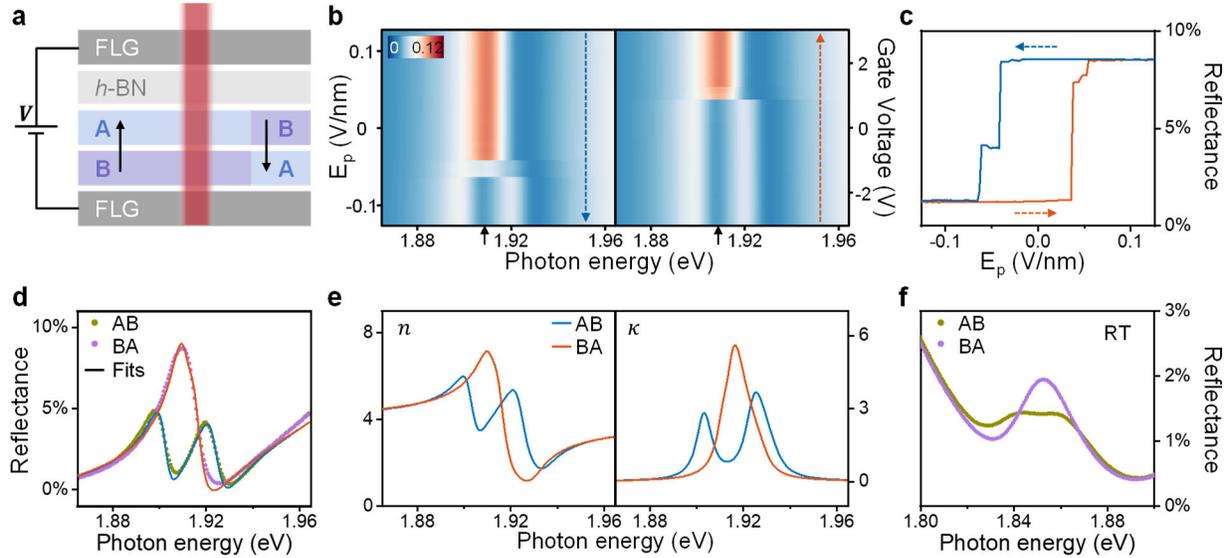

**Figure 3 | Non-volatile switching of excitonic resonance.** **(a)** Schematic of a 3R-MoS₂ bilayer screened asymmetrically by few-layer graphene (FLG) below and an h-BN layer above. The top FLG serves as an electrode to apply a vertical electric field. The bilayer MoS₂ has a preexisting domain wall (DW) that can be released when the electric field surpasses the pinning threshold. Black arrows indicate the direction of spontaneous polarization. **(b)** Local reflectance spectrum under a broadband white light source focused on a μm-sized spot, as a function of the poling field in both negative (left) and positive (right) scan directions. An electric poling field is applied for one second and then turned off, followed by a reflectance measurement. All measurements are performed at 4.5 K unless specified otherwise. **(c)** Reflectance at 1.909 eV as the poling field is scanned in negative (blue) and positive (red) directions. A hysteretic loop with a large reflectance contrast is observed and the origin of the intermediate steps is discussed in the main text. **(d, e)** The refractive index (n) and extinction coefficient ($\kappa$) for AB and BA stackings near the excitonic resonances are extracted by fitting the reflectance spectra. The maximum modulations in both optical constants are about 4. **(f)** At room temperature, the excitonic peak splitting is not resolvable due to thermal broadening, but the reflectance at the resonance peak can be switched between 1.42% and 1.95%, giving rise to a modulation depth of 27%.

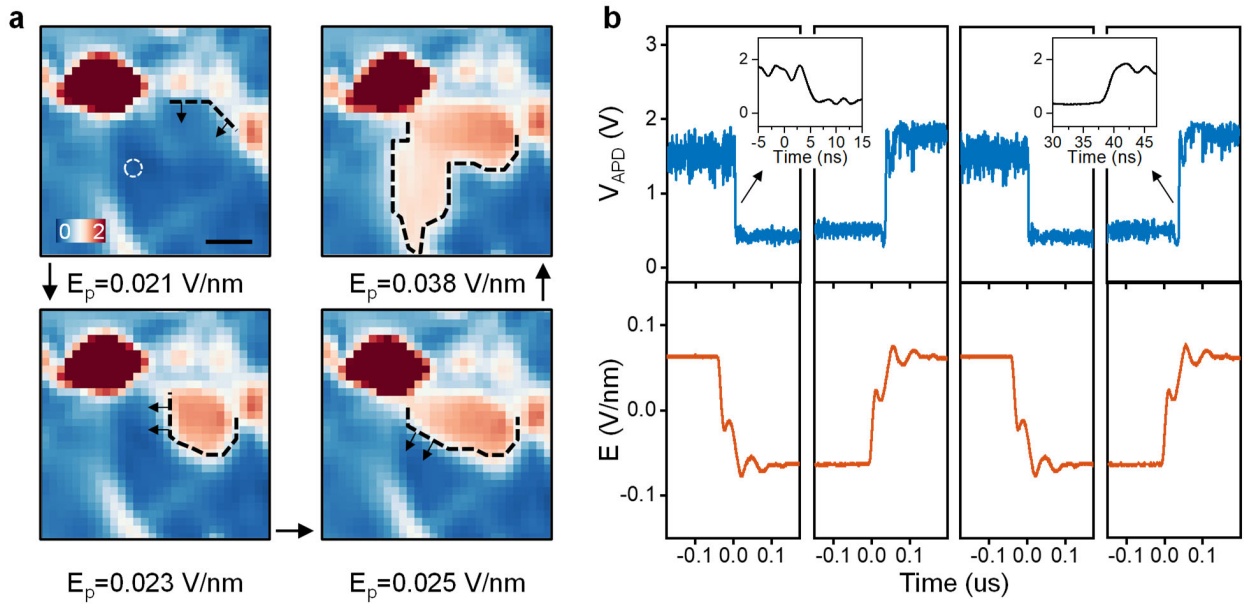

**Figure 4 | Nanosecond reflectance switching via domain wall propagation. (a)** Optical reflectance mapping (hν =1.904 eV) after different poling fields were applied shows that the DW was initially pinned at the upper right edge and propagated leftward and downward after release. The initial, intermediate, and final states are repeatable and non-volatile, as DW is trapped by the same local pinning centers. Other strongly reflective features in the map match the locations of the bubbles, wrinkles, and flake edges in the device. The circle marks the measurement spot for data presented in panel b and Fig. 3. Scale bar: 3 μm. **(b)** A clear correlation is observed in the time-domain between the optical reflectance and the applied electric field at both the fall and rise edges. The insets show a transient time of about 2.5 ns, limited by the measurement instruments. The abrupt step in the APD signal is significantly shorter than the electric field transition, confirming that the switching is caused by the sudden release and rapid propagation of the single domain as seen in panel a.

**Methods**

**Material and device fabrication:** The 3R-MoS$_2$ bilayer device featuring an asymmetric screening environment was fabricated using a layer-by-layer dry transfer technique under ambient conditions. Single crystals of 3R-MoS$_2$ were purchased from HQ Graphene.

**EFM characterization:** Electrostatic force microscopy (EFM) was conducted using a Molecular Vista atomic force microscope (AFM) in the tapping mode[41]. Gold-coated AFM tips (Tap300GB-G) with a first mechanical resonance of ~ 300 kHz, a second resonance of ~ 1500 kHz, and a force constant of ~ 40 N/m were used. For topographic information, the cantilever was driven by the piezoelectric actuator at its second resonance with an oscillation amplitude close to 2 nm. To measure the surface potential contrast, the AFM tip was grounded and an AC voltage oscillating at the fundamental resonance with an amplitude of 0.2 V was applied to the highly conductive Si substrate. The resulting electrostatic force was measured by the oscillation amplitude of the cantilever at its fundamental resonance.

**Optical spectroscopy:** Reflection spectroscopy was measured at the base temperature of either an Attodry-2100 (1.6 K) or CryoAdvance-50 (4.5 K) closed-cycle optical cryostat. A spatially filtered broadband tungsten halogen lamp was focused on a $\mu$m-sized spot on the device via a microscope objective (N.A. is 0.65 for Attodry-2100 and 0.5 for CryoAdvance-50). The reflected light was spectrally resolved using a spectrometer (Princeton Instruments) equipped with a thermoelectric-cooled CCD camera. The reflectance contrast spectra in the bare flake are normalized to the reference spectrum collected outside the flake. The reflectance spectra in the encapsulated device are normalized to the reference spectrum from a silver mirror.

**Switching speed measurement:** The switching speed was measured in the reflection geometry shown in Fig. S9. The emission from a temperature-controlled diode laser with a wavelength of 648.2 nm was focused onto a diffraction-limited spot of about 1.3-$\mu$m diameter on the device, which was cooled to the base temperature of CryoAdvance-50. The focus intensity was kept below 100 kW/cm$^2$ to avoid saturation. The reflected light intensity was detected by a silicon avalanche photodetector with a 400-MHz bandwidth (Thorlabs APD430A). The APD voltage was recorded by a 4-GHz digital oscilloscope (Tektronix MSO64B) using the single trigger mode. The square wave output by an arbitrary waveform generator (Stanford Research DS345) was used to switch the device and trigger the oscilloscope. The APD-oscilloscope combination provides a time resolution of about 2.5 ns, as reflected in the calibration of the instrument response function using a femtosecond laser (Fig. S10). The lower bound of the domain wall velocity was estimated from the ratio of the focus diameter to the instrument response time.